\documentclass[a4paper,11pt,onecolumn]{article}

\usepackage{amsmath}
\usepackage{verbatim}
\usepackage{dcolumn}
\usepackage{multirow}
\usepackage{hyperref}
\usepackage{lineno}
\usepackage{wasysym}
\usepackage{colortbl}
\usepackage{hyperref}

\usepackage{amssymb}
\usepackage{units}
\usepackage{subfigure}
\usepackage{color}
\usepackage{upgreek}
\usepackage{eepic, epic}
\usepackage[bf]{caption}       

\usepackage{graphicx}

\def\TTra{\textsuperscript{\texttrademark}}
\def\TReg{\textsuperscript{\textregistered}}

\begin{document}

\title{A Lightweight Field Cage for a Large TPC Prototype for the ILC}
\author{Ties Behnke, Klaus Dehmelt, Ralf Diener, Lea Hallermann, \\Takeshi Matsuda, Volker Prahl, Peter Schade
\vspace{0.7cm}\\
DESY, Hamburg, Germany}


\maketitle
\begin{abstract}
  We have developed and constructed the field cage of a prototype Time Projection Chamber   
  for research and development studies for a detector at the International Linear Collider.
  This prototype has an inner diameter of $\unit[72]{cm}$ and a length of $\unit[61]{cm}$.
  The design of the field cage wall was optimized for a low material budget of $1.21\%$ of a radiation length
  and a drift field homogeneity of $\Delta E/E \lesssim 10^{-4}$. 
  Since November 2008 the prototype has been part of a   
  comprehensive test beam setup at DESY and used as a test chamber for the development of
  Micro Pattern Gas Detector based readout devices.

\end{abstract}


\section{Introduction}
\label{introsection}
A Time Projection Chamber (TPC) is planned as the main tracking detector for the International Large Detector, ILD,  
a proposed detector for the International Linear Collider, ILC~\cite{:2007sg}.
This TPC will be confronted with multi-jet events with high track multiplicities. It has to provide
a very high tracking efficiency and precision while maintaining robustness towards machine backgrounds. 
 The detailed performance requirements for the ILD TPC are summarized in the ILD Letter of Intent~\cite{ILDLOI} 
and shown in Table~\ref{designgolasILDTPC}.

The momentum resolution goal is $\delta(1/p_\perp) \approx \unit[9\times 10^{-5}]{GeV^{-1}}$ for the TPC alone and derived from 
requirements on the physics performance of the ILD detector. This is directly linked with 
the point resolution of the TPC which should be better than $\unit[100]{\upmu m}$ in the $r\varphi$ plane, perpendicular to the beam pipe.
Of particular importance for the operation of the TPC will be the minimization of the material budget 
of the field cage structure. A low material budget is essential to suppress
conversion and multiple scattering processes before particles reach the calorimeter.

The performance goals significantly exceed the corresponding 
numbers reached by prior TPCs in collider experiments (e.g.~\cite{Alice, Anderson:2003, Atwood:1991bp}). 

During the last few years, Micro Pattern Gas Detector (MPGD) amplification systems were under 
study within the LCTPC collaboration \cite{LCTPC}
for the readout of the ILD TPC. The investigated MPGDs are 
Gas Electron Multiplier (GEM)~\cite{Sauli:1997qp} and Micromegas~\cite{Giomataris:1995fq} in combination with a pad or
pixel readout system. Both, GEMs or Micromegas devices can be mounted on a lightweight support    
and allow for the construction of a TPC end plate with a low material budget. In addition, they provide a flat and homogeneous surface
without large $\vec{E}\times \vec{B}$ effects in the vicinity of the readout plane. 

\begin{table}[t]
\centering
\begin{tabular}{ll}
\hline
Size                            & inner field cage $\varnothing$: $\unit[0.65]{m}$ \\
                                & outer field cage $\varnothing$: $\unit[3.6]{m}$ \\
                                & total length: $\unit[4.3]{m}$\\
point resolution in $r \varphi$ & $\sigma_\perp < \unit[100]{\upmu m}$ modulo $\varphi$ \\
point resolution in $z$         & $\sigma_z< \unit[0.5]{mm}$  modulo $\theta$\\
2-hit resolution in $r\varphi$  & $\sim \unit[2]{mm}$ (modulo track angles)\\
2-hit resolution in $z$         & $\sim \unit[6]{mm}$ (modulo track angles)\\
momentum res.                   & $\delta(1/p_\perp) \approx \unit[9\cdot 10^{-5}]{GeV^{-1}}$ \\
d$E$/d$x$ resolution            & $\sim \unit[5]{\%}$ \\
TPC material budget                 & $\lesssim \unit[0.01]{X_0}$ of the inner barrel \\
                                & $\lesssim \unit[0.04]{X_0}$ to the outer barrel \\
                                & $\lesssim \unit[0.15]{X_0}$ to the end caps \\
efficiency (TPC alone)          & $> \unit[97]{\%}$ (for $p_\perp > \unit[1]{GeV/c}$) \\
\hline
\end{tabular}
\caption{{\it{Design goals for the ILD TPC}}~\cite{ILDLOI}.}
\label{designgolasILDTPC}
\end{table}

First feasibility studies for a GEM or Micromegas based TPC readout were 
carried out by several research groups. The studied readout structures had sizes of typically 
$\unit[10]{cm} \times \unit[10]{cm}$ (e.g. \cite{ILDLOI} and references therein).

\begin{figure}[t!]
\centering
\includegraphics[width=0.7\textwidth]{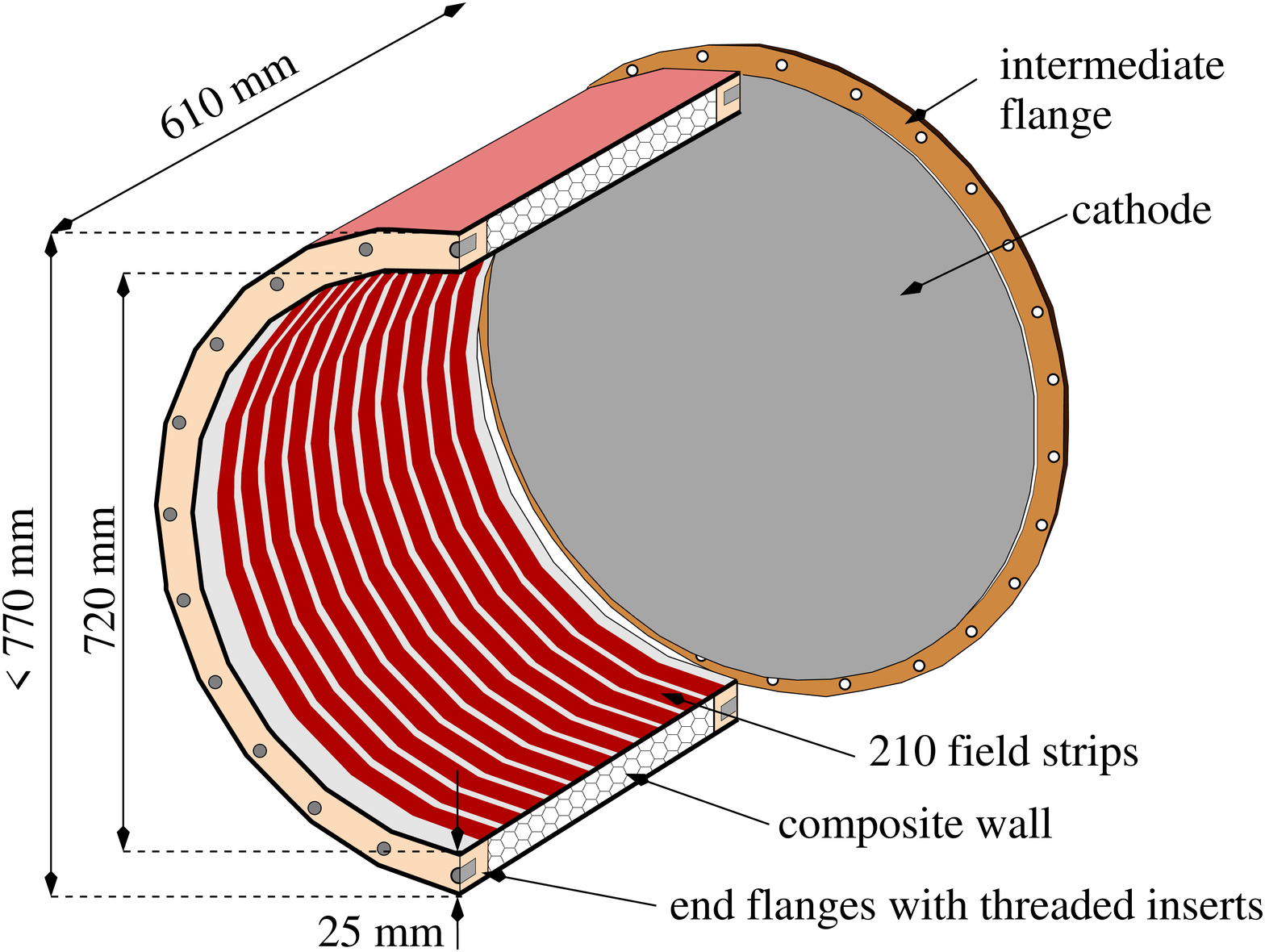}
\caption{\it{Overview of the design of the field cage:  
    Complementary to the field cage barrel, a cathode end plate was
    constructed. The cathode is supported inside the field cage by an intermediate flange.}}
\label{SketchofChamber}
\vspace{6ex}
\centering
\includegraphics[width=0.7\textwidth]{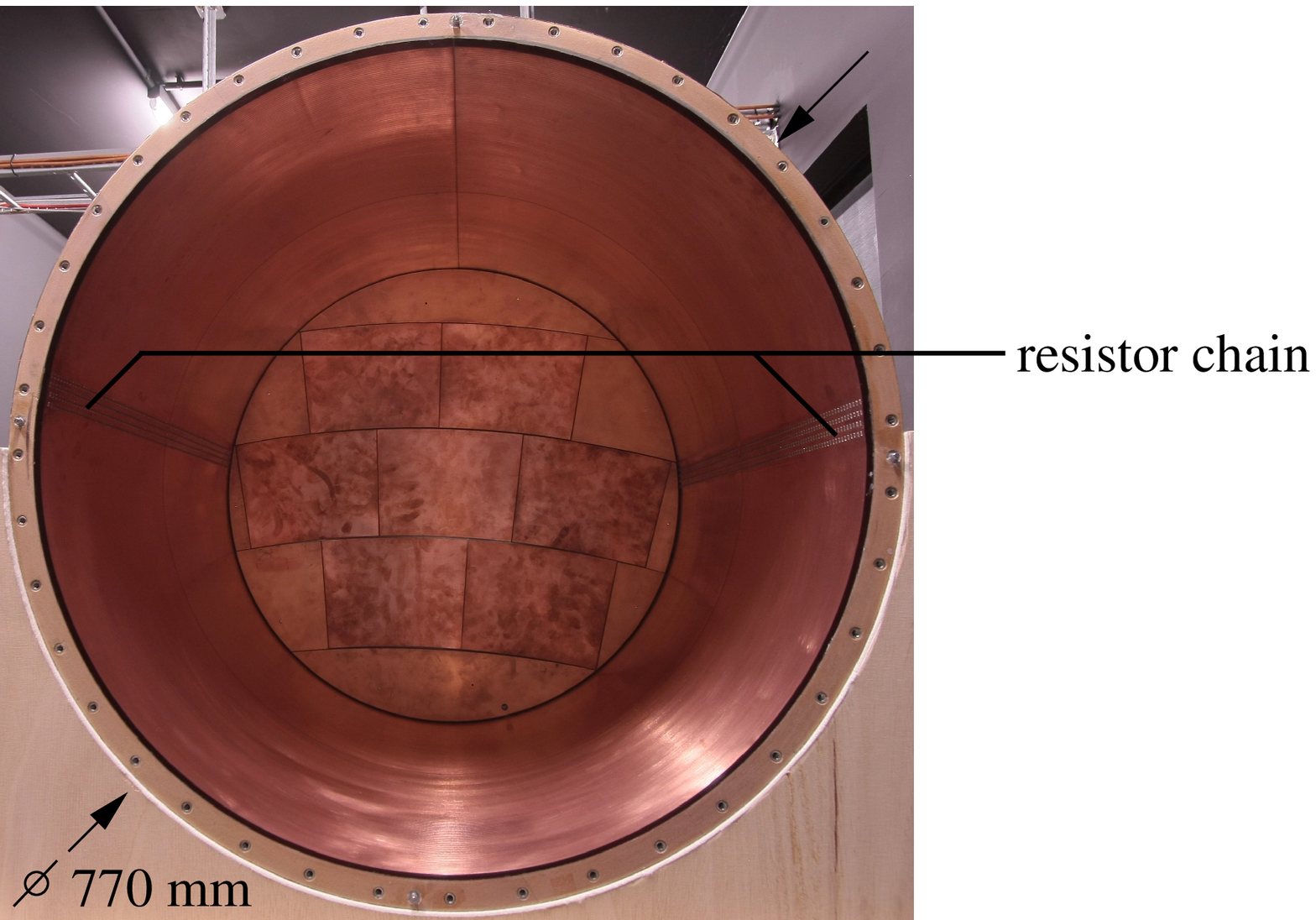}
\caption{{\it{View into the field cage from the cathode side: The anode is assembled with 
an end plate, which was constructed within the LCTPC collaboration}} \cite{LCTPC}. 
{\it{Two resistor chains are installed on the inside wall of the barrel and interconnect the field strips.}}}
\label{LPFertigpicture}
\end{figure}

The next step is to demonstrate a TPC with several prototype readout modules in a strong magnetic field.
A test beam infrastructure for the studies planned was realized at DESY in the framework of the
EUDET project \cite{EUDET}. The setup provides a superconducting solenoid magnet with a bore diameter of $\unit[85]{cm}$, 
a usable length of about $\unit[1]{m}$ and a magnetic field strength of up to $\unit[1.25]{T}$.
The TPC Prototype has an outer diameter of $\unit[77]{cm}$ and a length of $\unit[61]{cm}$ (Fig. \ref{SketchofChamber} and 
Fig.~\ref{LPFertigpicture})
and is dimensioned to be operated inside the magnet.

The diameter of this Large TPC Prototype (LP) is similar to the inner field cage of the ILD TPC. Moreover, the 
ratio $L/B$ of the TPC drift distance $L$ to the magnetic field strength $B$ is the same for the LP ($B=\unit[1]{T}$, $L=\unit[60]{cm}$) 
and the ILD TPC ($B = \unit[3.5]{T}$, $L=\unit[215]{cm}$). 
If this ratio remains constant, the magnitude of acceptable electric field inhomogeneities inside the TPC drift volume will
also remain the same.
Therefore, the relative mechanical accuracy specifications are similar for the LP and the ILD TPC.

In the following, optimization studies for the LP field cage and its construction are discussed.
Based on the experience gained with the LP a preliminary design for the ILD TPC field cage wall is 
proposed.

\section{Requirements for the Field Cage}
The design of the Large TPC Prototype was optimized towards a low material budget of the walls, 
a high homogeneity of the electric drift field and an adequate maximum operational voltage.

The material budget per wall of the barrel was required to be close to the design goal 
of $\unit[1\%]{X_0}$ for the ILD TPC.

Radial components $\Delta E_\text{r}$ of the electric drift field inside the LP volume should
not exceed $\Delta E_\text{r}/E \lesssim 10^{-4}$. 
This limits systematic effects on the resolution due to field inhomogeneities to less than $\unit[30]{\upmu m}$. 
Controlling the field distortions on a level of $10^{-4}$ requires a mechanical accuracy of the field cage
in the 100-$\upmu$m regime.

The LP has to allow for operations with various gases with an overpressure of up to $\unit[10]{mbar}$.  
Deformations of the structure due to the overpressure should stay below $\unit[100]{\upmu m}$. 
The anticipated maximum drift fields are in the range of $\unit[350]{V/cm}$, which require long term operations  
without voltage breakdowns with $\unit[20]{kV}$ permanently applied to the cathode of the LP.

\section{Design of the Wall Structure}
\label{BendingtestChapter}
\begin{figure}[t!]
\centering
\subfigure[honeycomb material]{\includegraphics[height = 0.48\textwidth]{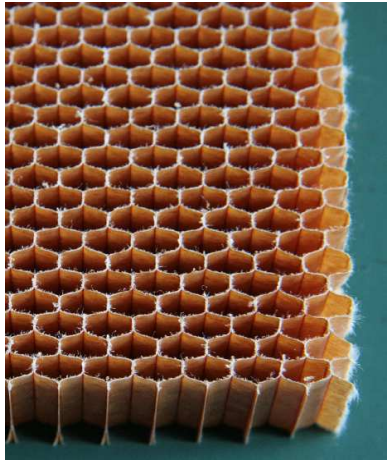}
\label{Honeycombsample}
}
\hspace{1cm}
\subfigure[wall sample]{\includegraphics[height = 0.48\textwidth]{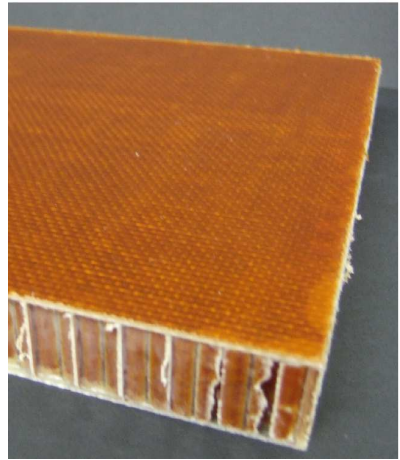}\label{Wallsample}
}
 \caption{\it{Composite wall structure: (a) Over-expanded honeycomb was used for the construction of the LP wall. 
     The cells of this material are expanded in one direction and have an almost rectangular shape. 
     The modified cell structure increases the flexibility of the material perpendicular to the direction of the expansion 
     and allows for the construction of cylindrical structures. 
     (b) Sample piece of the wall, as used for mechanical and electrical tests with $\unit[400]{\upmu m}$ thick GRP layers.}}
\label{Samplepieces}
\end{figure}

The field cage barrel of the LP was built as a lightweight sandwich structure. The wall consists of a 
23.5-mm thick over-expanded aramid honeycomb material (Fig.~\ref{Honeycombsample}) which is embedded 
between two layers of glass-fiber reinforced plastic (GRP) and a polyimide layer for electrical insulation.

A low material budget of the wall was achieved by minimizing the thickness of the GRP layers. 
The wall was tested for mechanical robustness and high voltage stability.

\begin{figure}[t]
  \centering
  \subfigure[four-point bending test setup]{\includegraphics[width=0.48\textwidth]{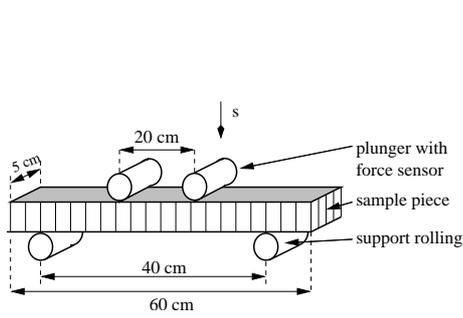}
    \label{bendingtestsetup}} \hfill
  \subfigure[test results for two sample pieces]{\includegraphics[width=0.48\textwidth]{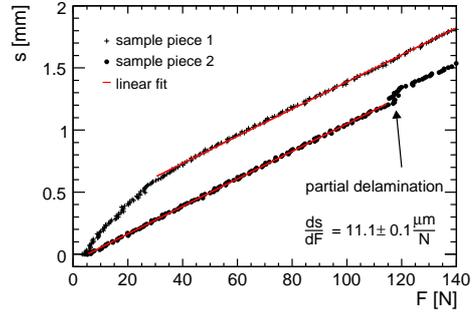}
    \label{testresult}}
  \caption{\it{Four-point bending test: (a)
      In the test setup, the pieces rest on two rollings with a distance
      of $\unit[40]{cm}$ while two similar rollings in a distance of $\unit[20]{cm}$ press centrally against the sample. 
      (b) The applied force $F$ and the elongation $s$ are measured in parallel. The dependence $s(F)$ is 
      linear with an equal slope for both samples. In case of the first sample the linear range starts only at forces
      of about $40$\,N due to an improper preparation of the measurement apparatus.
      The second sample suffers from first damage at forces of about 120\,N (partial delamination).
    }}
\label{Bendingtests}
\end{figure}

\subsection{Mechanical Robustness}
To test the mechanical properties of the field cage wall, several sample pieces were 
produced (Fig.~\ref{Samplepieces}).
Two sample pieces were subjected to a four point bending test 
(Fig.~\ref{Bendingtests})\footnote{The tests were performed in cooperation with the Technical University of Hamburg-Harburg.}.
For small forces $F$, the observed bending $s$ rises linear with the applied force $F$ according to 
\begin{align*}
\frac{\text{d}s}{\text{d}F} = \unit[11.1 \pm 0.1]{\frac{\upmu m}{N}} \quad (F<\unit[100]{N}).
\end{align*} 
To limit the deflection $s$ to below $\unit[100]{\upmu m}$, the force $F$ on the structure must not exceed $\unit[10]{N}$.
This corresponds to a maximum pressure of $\unit[5]{mbar}$ on the sample.
At larger forces the samples suffer from partial delamination and are irreversibly damaged. 

Translated from the flat geometry of the test setup to the cylindrical structure of the LP, the bending of the barrel is 
reduced by a factor of approximately 80. The factor was determined in FEM calculations. 
To keep the wall deflection below $\unit[100]{\upmu m}$, 
the overpressure inside the LP should not exceed $\unit[400]{mbar}$. 
Thus, the field cage barrel is mechanically robust for operations at the envisaged overpressure of $\unit[10]{mbar}$.

\subsection{High Voltage Stability}
\label{ElectrostaticDesign}
\begin{figure}[t!]
 \centering
\subfigure[displaced mirror strips, lying on the intermediate potential of the two adjacent field strips]{
\begin{tabular}{c}
\includegraphics[width=0.44\textwidth]{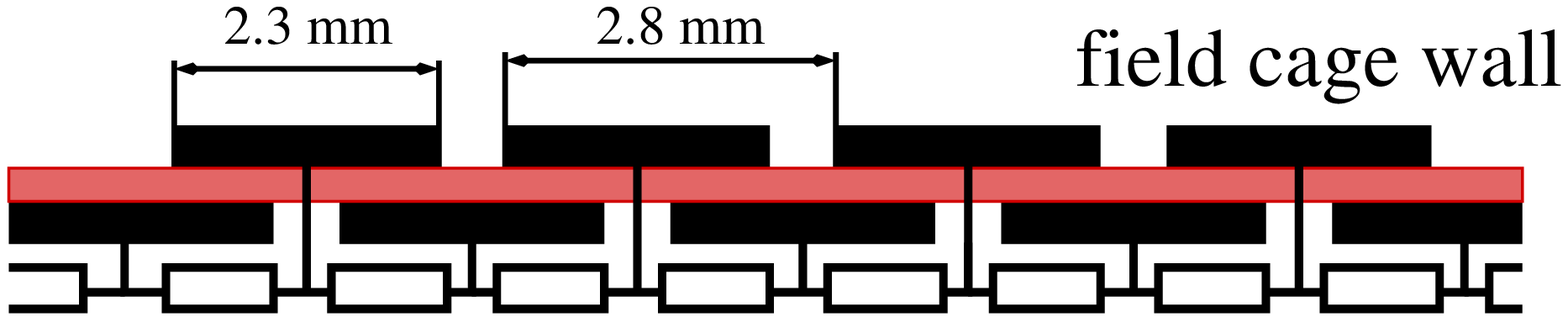}\\
\includegraphics[width=0.44\textwidth]{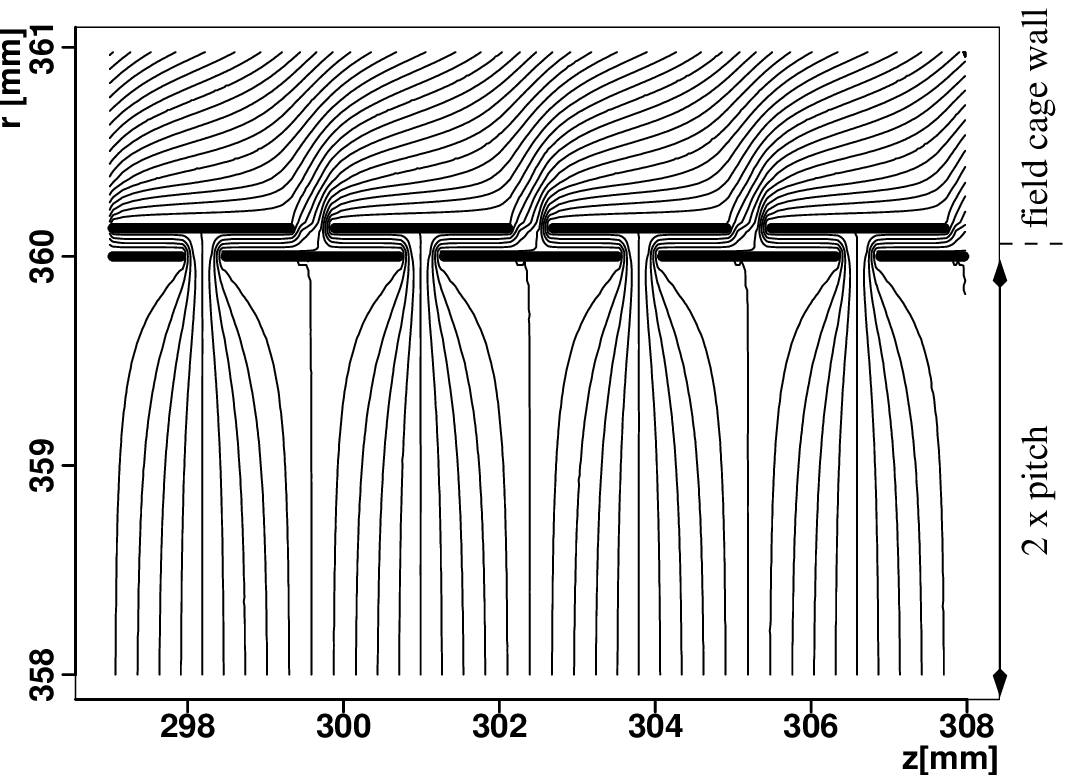}
\end{tabular}
\label{standardStreifen}
}
\subfigure[large mirror strips, directly connected to the field strips]{
\begin{tabular}{c}
\includegraphics[width=0.44\textwidth]{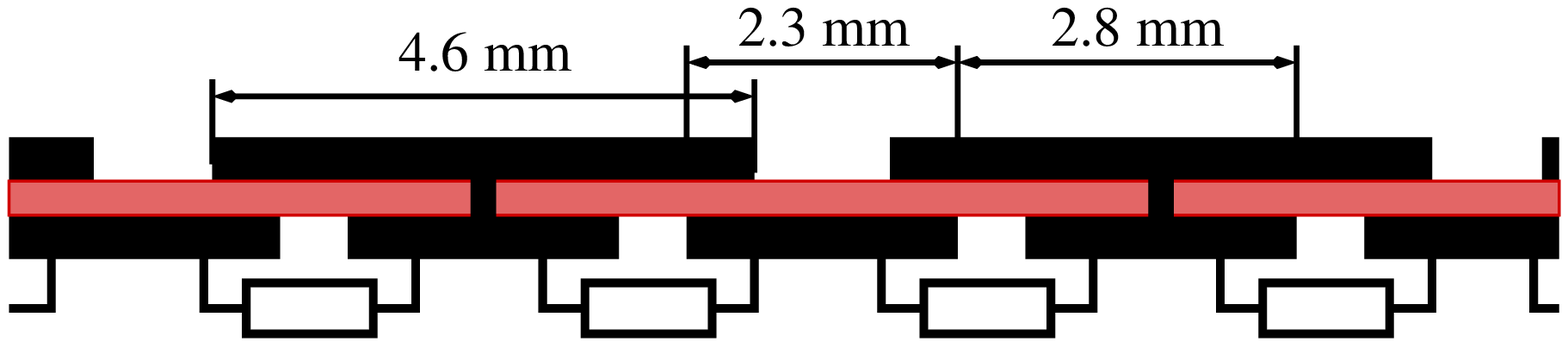}\\
\includegraphics[width=0.44\textwidth]{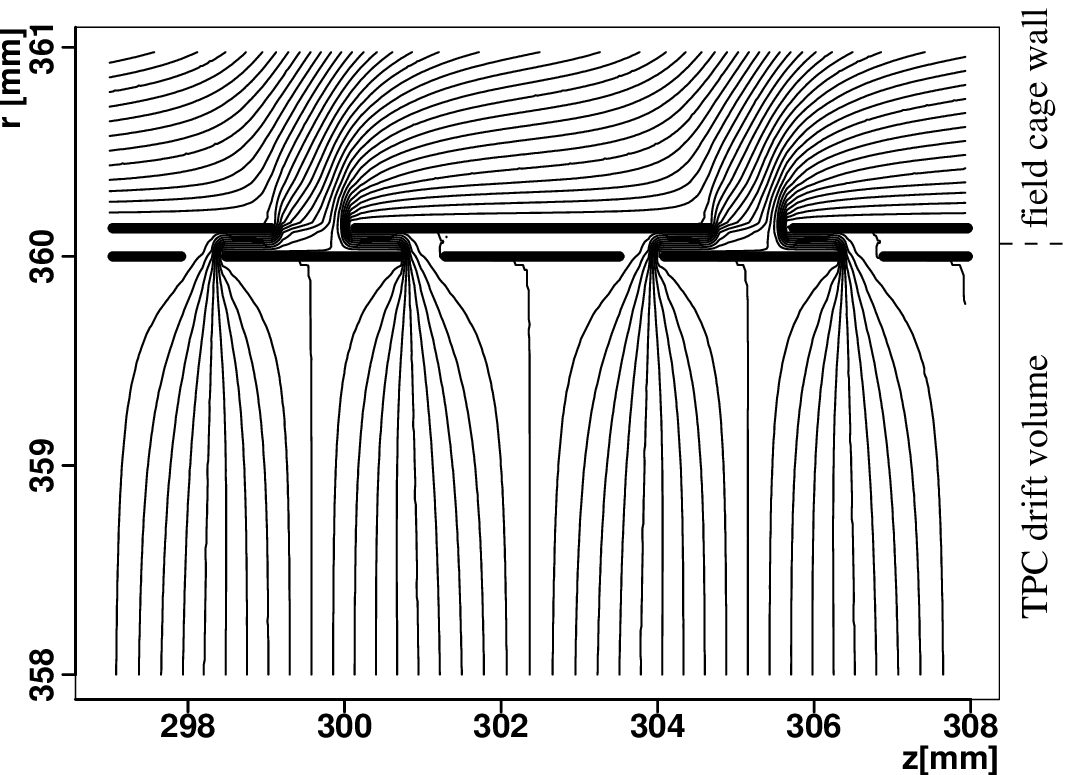}
\end{tabular}
\label{breiteParallel}
} \hfill
\caption{\it{Calculated electric equipotential lines on the inner wall of the field cage: 
  (a) A standard layout with displaced mirror strips 
  covering the gaps between the field strips. (b) A layout with extended mirror strips.}}
\end{figure}
To guarantee the operational safety, high-voltage breakdown tests were performed. 
For this purpose, the wall samples were installed in air between a parallel plate capacitor and $\unit[30]{kV}$ applied
for $\unit[24]{h}$. 

The samples evaluated contained polyimide insulation layers with thicknesses between $\unit[50]{\upmu m}$ 
and $\unit[150]{\upmu m}$. No breakdowns were observed. 
 
The final design of the LP wall contains a polyimide insulation layer of $\unit[125]{\upmu m}$ thickness and the LP is expected
to be high-voltage stable for long term operations with voltages of $\unit[20]{kV}$.

\section{Design of the Field Forming Elements}
\begin{figure}[t!]
\includegraphics[width=0.9\textwidth]{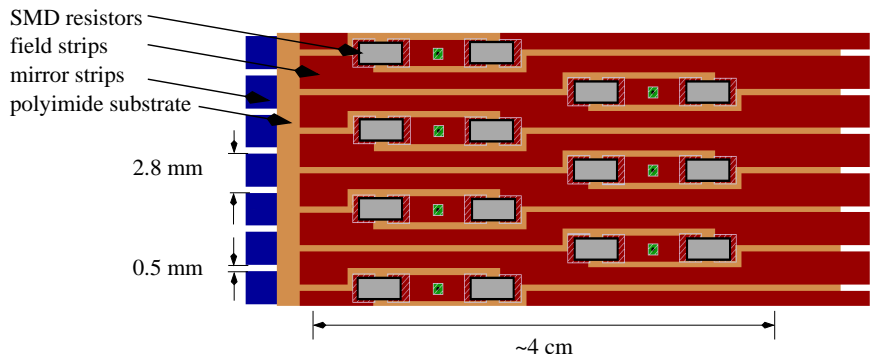}
\caption{\it{Layout of the resistor chains on the field strip boards for the LP: 
Two neighboring strips are connected by two surface mount (SMD) resistors via an intermediate connection which tabs through the board 
to the mirror strip. This corresponds to the strip design shown in Figure~\ref{standardStreifen}.
}}
\label{stripLayout}
\end{figure}

The inside of the LP barrel is covered with conductive copper rings (see Fig.~\ref{SketchofChamber} and Fig.~\ref{LPFertigpicture}). 
These field shaping strips lie on stepwise decreasing potentials from the anode to the cathode and define 
the boundary condition for the electric field along the inside of the TPC barrel. 
A second layer, the mirror strips, is installed directly under the field strips. Each mirror strip covers 
the gap between two field strips in front. 
Together, the two layers provide a shielding against external electrical influences on the internal field.

With the help of finite-element field calculations several strip arrangements were investigated.
The layout chosen for the LP (Fig.~\ref{standardStreifen}) is a typical arrangement used in TPCs (e.g.~\cite{Bowdery:321763}).
The field shaping strips have a pitch of $\unit[2.8]{mm}$ and are intersected by $\unit[0.5]{mm}$ gaps, 
while the mirror strips are a copy of the field strips but displaced by half the pitch. Each mirror strip lies on the 
intermediate potential of the two adjacent field strips. These potentials are applied by a resistor chain. 
If the insulation layer between the field strips and the mirror strips is kept thin compared to the strip's width,
field distortions occur only in a narrow band with a thickness of
two times the pitch along the inner field cage wall.

A second design was evaluated as an alternative (Fig.~\ref{breiteParallel}).
Here, only every second field strip is connected to a mirror strip while each mirror strip covers two
gaps. As a result, the drift field becomes homogeneous at a distance of three times the pitch from the wall. 
This arrangement would allow for a simpler design of the resistor chain. 

In the LP, the strip design is realized on a $\unit[61]{cm} \times \unit[226]{cm}$ large flexible printed circuit board -- 
the width and length of the board correspond to the length and inner circumference of the field cage, respectively.
The board consists of a 75-$\upmu$m thick polyimide carrier foil with 35-$\upmu$m thick copper 
field and mirror strips on either side, respectively.
The side with the field strips accommodates places to solder surface-mount resistors (Fig.~\ref{stripLayout}). 
Two of these resistor chains are installed on the inside wall of the field cage, in diametrical opposite positions 
(see Fig.~\ref{LPFertigpicture}).

For technical reasons, the final 61-cm wide board was split up into two pieces. 
These two half-boards were produced by industry\footnote{Optiprint, Innovative PCB Solutions, \href{http://www.optiprint.ch}{http:/\hspace{-0.6mm}/www.optiprint.ch}} and 
afterwards combined into one piece.

The field strip board was assembled with resistors and electrically tested prior to the construction of the field cage. 
It is equipped with $\unit[1]{M\Omega}$ resistors with a measured spread of $\Delta R \lesssim \unit[100]{\Omega}$, or 
$\Delta R/R \lesssim 10^{-4}$.
The installation of the field strip board into the field cage is described in Section~\ref{ProductionSection}.

\section{Cross Section of the Field Cage Wall}
\label{wallcrosssectionsubsection}
\begin{figure}[t]
\centering
\includegraphics[width=0.75\textwidth]{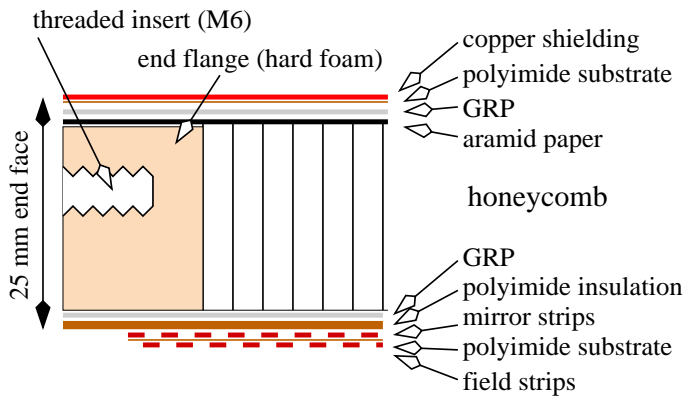}
\caption{\it{Cross section of the Large Prototype field cage wall.}}
\label{SketchofWall}
\end{figure}
\begin{table}
\centering
\begin{tabular}{ll}
\hline
insulation layer      & DuPont\TTra, Kapton\TReg\,500HN \\
aramid honeycomb      & Hexel, HexWeb\TReg\\
                      & HRH 10/OX-3/16-1.8\\
hard foam end flanges & SP, Corecell\TTra\, S-Foam \\
aramid paper          & DuPont\TTra, Nomex\TReg\,410\\
\hline
\end{tabular}
\caption{\it{Materials used for the construction of the field cage.}}
\label{materials}
\end{table}

The wall of the field cage consists of four main components. 
Figure~\ref{SketchofWall} displays the cross section in detail 
and Table~\ref{materials} summarizes the materials used in the wall laminate.

An electrical shielding layer on the outside of the barrel is realized by a layer of $\unit[10]{\upmu m}$ thick copper 
on a polyimide carrier of $\unit[50]{\upmu m}$ thickness.
The copper layer is electrically grounded and confines the electric field of the TPC
to the inside of the field cage.

The bulk of the wall consists of the honeycomb spacer material sandwiched between two GRP layers. 
The honeycomb is $\unit[23.5]{mm}$ thick and has a density of $\unit[29]{kg/m^3}$. 
A layer of aramid paper was introduced on the outside of the honeycomb for constructional reasons 
(see Sec.~\ref{ProductionSection}).

\begin{table}[t!]
\centering
\begin{tabular}{lrrr}
layer of the wall       &  $\unit[d]{[cm]}$  & $\unit[X_\text{0}]{[cm]}$ & $\unit[d / X_\text{0}]{[\%]}$\\
\hline
\hline
copper shielding        &  0.001               &                   1.45   & 0.07 \\
polyimide substrate     &  0.005               &                  32.65   & 0.02 \\ 
outer GRP               &  0.03                &                  15.79   & 0.19 \\ 
aramid paper            &  0.007              &                   29.6   & 0.02 \\
honeycomb               &  2.35               &                   1383   & 0.17 \\
inner GRP               &  0.03                &                  15.79   & 0.19 \\ 
polyimide insulation    &  0.0125              &                  32.65   & 0.04 \\
mirror strips           &  0.8 $\cdot$ 0.0035  &                   1.45   & 0.19 \\
polyimide substrate     &  0.0050              &                  32.65   & 0.02 \\ 
field strips            &  0.8 $\cdot$ 0.0035  &                   1.45   & 0.19 \\
                        &                     &                          &      \\
epoxy glue              &  $\approx 6 \cdot 0.007$    &         $\approx 35.2$   & 0.12 \\
\hline
\hline
& & $\sum$& 1.21\\
\end{tabular}
\caption{{\it{Composition and radiation lengths of the materials in the field cage wall: the thickness of the different 
layers were derived from the specifications of the used materials. 
Material densities and radiation lengths were taken from}}~\cite{pdgbooklet}.
\it{The thickness of the copper layers are reduced by factors of 0.8 because the field strips cover only $80\%$ 
of the inner field cage barrel.}}
\label{radlenghttable}
\end{table}

A 125-$\upmu$m thick polyimide layer ensures the high-voltage stability of the wall. 
This polyimide layer alone has a breakdown voltage of about $\unit[20]{kV}$. The 
honeycomb sandwich is non conductive and contributes further to the high voltage stability of the wall laminate.

The field and mirror strips, on the inside of the barrel 
(see Fig.~\ref{standardStreifen}) suppress the influence of the ground potential of the outer shielding
on the drift field and guarantee an electric field homogeneity of $\Delta E/E \lesssim 10^{-4}$.

The field cage wall is terminated on the anode and cathode side by end flanges made of hard foam (see Tab.~\ref{materials}).
These flanges have a height of $\unit[23.5]{mm}$, which matches the height of the honeycomb material, and 
are populated with threaded stainless steel inserts for the attachment of the anode and cathode end plates.

The radiation length of the wall is
\begin{align*}
X^\text{wall} = \unit[1.21 \pm 0.10]{\%} X_\text{0}.
\end{align*}

In the calculation  of $X^\text{wall}$ (Tab.~\ref{radlenghttable}), GRP was assumed to consist of $2/3$ glass fiber 
and $1/3$ epoxy glue. In addition, the  
thickness of the epoxy layers used to glue together the different layers of the wall was estimated to be $\unit[70 \pm 30]{\upmu m}$ 
thick each.

\section{Specification  of  Mechanical Accuracy}
 \begin{figure}[t]
 \centering

\setlength{\unitlength}{2800sp}%
\begingroup\makeatletter\ifx\SetFigFont\undefined%
\gdef\SetFigFont#1#2#3#4#5{%
  \reset@font\fontsize{#1}{#2pt}%
  \fontfamily{#3}\fontseries{#4}\fontshape{#5}%
  \selectfont}%
\fi\endgroup%
\begin{picture}(4858,4056)(526,-5323)
\thinlines
{\color[rgb]{0,0,0}\put(1801,-1936){\line( 1, 0){3000}}
}%
{\color[rgb]{0,0,0}\put(1801,-2236){\line( 1, 0){3000}}
}%
{\color[rgb]{0,0,0}\put(1951,-2236){\line( 0, 1){300}}
}%
{\color[rgb]{0,0,0}\put(1876,-2236){\line( 0, 1){300}}
}%
{\color[rgb]{0,0,0}\put(2026,-2236){\line( 0, 1){300}}
}%
{\color[rgb]{0,0,0}\put(2101,-2236){\line( 0, 1){300}}
}%
{\color[rgb]{0,0,0}\put(2176,-2236){\line( 0, 1){300}}
}%
{\color[rgb]{0,0,0}\put(2251,-2236){\line( 0, 1){300}}
}%
{\color[rgb]{0,0,0}\put(2326,-2236){\line( 0, 1){300}}
}%
{\color[rgb]{0,0,0}\put(2476,-2236){\line( 0, 1){300}}
}%
{\color[rgb]{0,0,0}\put(2401,-2236){\line( 0, 1){300}}
}%
{\color[rgb]{0,0,0}\put(2551,-2236){\line( 0, 1){300}}
}%
{\color[rgb]{0,0,0}\put(2626,-2236){\line( 0, 1){300}}
}%
{\color[rgb]{0,0,0}\put(2701,-2236){\line( 0, 1){300}}
}%
{\color[rgb]{0,0,0}\put(3001,-2236){\line( 0, 1){300}}
}%
{\color[rgb]{0,0,0}\put(3076,-2236){\line( 0, 1){300}}
}%
{\color[rgb]{0,0,0}\put(3151,-2236){\line( 0, 1){300}}
}%
{\color[rgb]{0,0,0}\put(3226,-2236){\line( 0, 1){300}}
}%
{\color[rgb]{0,0,0}\put(3301,-2236){\line( 0, 1){300}}
}%
{\color[rgb]{0,0,0}\put(3376,-2236){\line( 0, 1){300}}
}%
{\color[rgb]{0,0,0}\put(3451,-2236){\line( 0, 1){300}}
}%
{\color[rgb]{0,0,0}\put(3526,-2236){\line( 0, 1){300}}
}%
{\color[rgb]{0,0,0}\put(3601,-2236){\line( 0, 1){300}}
}%
{\color[rgb]{0,0,0}\put(3676,-2236){\line( 0, 1){300}}
}%
{\color[rgb]{0,0,0}\put(3751,-2236){\line( 0, 1){300}}
}%
{\color[rgb]{0,0,0}\put(3826,-2236){\line( 0, 1){300}}
}%
{\color[rgb]{0,0,0}\put(3901,-2236){\line( 0, 1){300}}
}%
{\color[rgb]{0,0,0}\put(3976,-2236){\line( 0, 1){300}}
}%
{\color[rgb]{0,0,0}\put(4051,-2236){\line( 0, 1){300}}
}%
{\color[rgb]{0,0,0}\put(4126,-2236){\line( 0, 1){300}}
}%
{\color[rgb]{0,0,0}\put(4201,-2236){\line( 0, 1){300}}
}%
{\color[rgb]{0,0,0}\put(4276,-2236){\line( 0, 1){300}}
}%
{\color[rgb]{0,0,0}\put(4351,-2236){\line( 0, 1){300}}
}%
{\color[rgb]{0,0,0}\put(4426,-2236){\line( 0, 1){300}}
}%
{\color[rgb]{0,0,0}\put(4501,-2236){\line( 0, 1){300}}
}%
{\color[rgb]{0,0,0}\put(4576,-2236){\line( 0, 1){300}}
}%
{\color[rgb]{0,0,0}\put(4651,-2236){\line( 0, 1){300}}
}%
{\color[rgb]{0,0,0}\put(4726,-2236){\line( 0, 1){300}}
}%
{\color[rgb]{0,0,0}\put(2776,-2236){\line( 0, 1){300}}
}%
{\color[rgb]{0,0,0}\put(2851,-2236){\line( 0, 1){300}}
}%
{\color[rgb]{0,0,0}\put(2926,-2236){\line( 0, 1){300}}
}%
{\color[rgb]{0,0,0}\put(1801,-4636){\line( 1, 0){3000}}
}%
{\color[rgb]{0,0,0}\put(1801,-4936){\line( 1, 0){3000}}
}%
{\color[rgb]{0,0,0}\put(1951,-4936){\line( 0, 1){300}}
}%
{\color[rgb]{0,0,0}\put(1876,-4936){\line( 0, 1){300}}
}%
{\color[rgb]{0,0,0}\put(2026,-4936){\line( 0, 1){300}}
}%
{\color[rgb]{0,0,0}\put(2101,-4936){\line( 0, 1){300}}
}%
{\color[rgb]{0,0,0}\put(2176,-4936){\line( 0, 1){300}}
}%
{\color[rgb]{0,0,0}\put(2251,-4936){\line( 0, 1){300}}
}%
{\color[rgb]{0,0,0}\put(2326,-4936){\line( 0, 1){300}}
}%
{\color[rgb]{0,0,0}\put(2476,-4936){\line( 0, 1){300}}
}%
{\color[rgb]{0,0,0}\put(2401,-4936){\line( 0, 1){300}}
}%
{\color[rgb]{0,0,0}\put(2551,-4936){\line( 0, 1){300}}
}%
{\color[rgb]{0,0,0}\put(2626,-4936){\line( 0, 1){300}}
}%
{\color[rgb]{0,0,0}\put(2701,-4936){\line( 0, 1){300}}
}%
{\color[rgb]{0,0,0}\put(3001,-4936){\line( 0, 1){300}}
}%
{\color[rgb]{0,0,0}\put(3076,-4936){\line( 0, 1){300}}
}%
{\color[rgb]{0,0,0}\put(3151,-4936){\line( 0, 1){300}}
}%
{\color[rgb]{0,0,0}\put(3226,-4936){\line( 0, 1){300}}
}%
{\color[rgb]{0,0,0}\put(3301,-4936){\line( 0, 1){300}}
}%
{\color[rgb]{0,0,0}\put(3376,-4936){\line( 0, 1){300}}
}%
{\color[rgb]{0,0,0}\put(3451,-4936){\line( 0, 1){300}}
}%
{\color[rgb]{0,0,0}\put(3526,-4936){\line( 0, 1){300}}
}%
{\color[rgb]{0,0,0}\put(3601,-4936){\line( 0, 1){300}}
}%
{\color[rgb]{0,0,0}\put(3676,-4936){\line( 0, 1){300}}
}%
{\color[rgb]{0,0,0}\put(3751,-4936){\line( 0, 1){300}}
}%
{\color[rgb]{0,0,0}\put(3826,-4936){\line( 0, 1){300}}
}%
{\color[rgb]{0,0,0}\put(3901,-4936){\line( 0, 1){300}}
}%
{\color[rgb]{0,0,0}\put(3976,-4936){\line( 0, 1){300}}
}%
{\color[rgb]{0,0,0}\put(4051,-4936){\line( 0, 1){300}}
}%
{\color[rgb]{0,0,0}\put(4126,-4936){\line( 0, 1){300}}
}%
{\color[rgb]{0,0,0}\put(4201,-4936){\line( 0, 1){300}}
}%
{\color[rgb]{0,0,0}\put(4276,-4936){\line( 0, 1){300}}
}%
{\color[rgb]{0,0,0}\put(4351,-4936){\line( 0, 1){300}}
}%
{\color[rgb]{0,0,0}\put(4426,-4936){\line( 0, 1){300}}
}%
{\color[rgb]{0,0,0}\put(4501,-4936){\line( 0, 1){300}}
}%
{\color[rgb]{0,0,0}\put(4576,-4936){\line( 0, 1){300}}
}%
{\color[rgb]{0,0,0}\put(4651,-4936){\line( 0, 1){300}}
}%
{\color[rgb]{0,0,0}\put(4726,-4936){\line( 0, 1){300}}
}%
{\color[rgb]{0,0,0}\put(2776,-4936){\line( 0, 1){300}}
}%
{\color[rgb]{0,0,0}\put(2851,-4936){\line( 0, 1){300}}
}%
{\color[rgb]{0,0,0}\put(2926,-4936){\line( 0, 1){300}}
}%
{\color[rgb]{0,0,0}\put(1951,-3436){\line( 1, 0){375}}
}%
{\color[rgb]{0,0,0}\put(2521,-3443){\line( 1, 0){ 16}}
}%
{\color[rgb]{0,0,0}\put(2701,-3429){\line( 1, 0){375}}
}%
{\color[rgb]{0,0,0}\put(3271,-3436){\line( 1, 0){ 16}}
}%
{\color[rgb]{0,0,0}\put(3451,-3429){\line( 1, 0){375}}
}%
{\color[rgb]{0,0,0}\put(4021,-3436){\line( 1, 0){ 16}}
}%
{\color[rgb]{0,0,0}\put(1201,-3429){\line( 1, 0){375}}
}%
{\color[rgb]{0,0,0}\put(1771,-3436){\line( 1, 0){ 16}}
}%
{\color[rgb]{0,0,0}\put(4201,-3429){\line( 1, 0){375}}
}%
{\color[rgb]{0,0,0}\put(4771,-3436){\line( 1, 0){ 16}}
}%
{\color[rgb]{0,0,0}\put(4904,-3435){\oval(74,300)}
}%
{\color[rgb]{0,0,0}\put(1571,-3437){\oval(74,300)}
}%
{\color[rgb]{0,0,0}\put(1811,-5236){\line( 1, 0){2990}}
}%
{\color[rgb]{0,0,0}\put(1576,-3586){\line( 1, 0){3330}}
}%
{\color[rgb]{0,0,0}\put(1576,-3287){\line( 1, 0){3330}}
}%
{\color[rgb]{0,0,0}\put(1425,-5089){\line( 0,-1){150}}
\put(1425,-5239){\line( 1, 0){150}}
}%
{\color[rgb]{0,0,0}\put(1279,-4784){\line( 1, 0){296}}
}%
{\color[rgb]{0,0,0}\put(4880,-5049){\line( 1, 0){150}}
}%
{\color[rgb]{0,0,0}\multiput(4807,-4977)(6.25000,-6.25000){13}{\makebox(1.6667,11.6667){\SetFigFont{5}{6}{\rmdefault}{\mddefault}{\updefault}.}}
\multiput(4882,-5052)(-6.25000,-6.25000){13}{\makebox(1.6667,11.6667){\SetFigFont{5}{6}{\rmdefault}{\mddefault}{\updefault}.}}
}%
{\color[rgb]{0,0,0}\multiput(2101,-2311)(380.76923,0.00000){7}{\line( 1, 0){190.385}}
}%
{\color[rgb]{0,0,0}\multiput(2101,-4561)(380.76923,0.00000){7}{\line( 1, 0){190.385}}
}%
{\color[rgb]{0,0,0}\put(1651,-2836){\vector( 0,-1){450}}
}%
{\color[rgb]{0,0,0}\put(1651,-3811){\vector( 0, 1){225}}
}%
{\color[rgb]{0,0,0}\put(1279,-4185){\line( 1, 0){296}}
}%
{\color[rgb]{0,0,0}\put(5026,-5161){\framebox(300,300){}}
}%
{\color[rgb]{0,0,0}\put(1276,-5081){\framebox(300,1270){}}
}%
{\color[rgb]{0,0,0}\put(826,-2386){\line( 1, 0){1500}}
}%
{\color[rgb]{0,0,0}\put(1604,-2387){\vector( 1, 0){225}}
}%
{\color[rgb]{0,0,0}\put(2286,-2387){\vector(-1, 0){225}}
}%
{\color[rgb]{0,0,0}\put(2100,-2332){\line( 0,-1){ 55}}
}%
\put(5070,-5108){\makebox(0,0)[lb]{\smash{{\SetFigFont{12}{14.4}{\rmdefault}{\mddefault}{\updefault}{\color[rgb]{0,0,0}A}%
}}}}
\put(1501,-4111){\rotatebox{90.0}{\makebox(0,0)[lb]{\smash{{\SetFigFont{12}{14.4}{\rmdefault}{\mddefault}{\updefault}{\color[rgb]{0,0,0}A}%
}}}}}
\put(1000,-3361){\makebox(0,0)[lb]{\smash{{\SetFigFont{12}{14.4}{\rmdefault}{\mddefault}{\updefault}{\color[rgb]{0,0,0}axis}%
}}}}
\put(1576,-3136){\rotatebox{90.0}{\makebox(0,0)[lb]{\smash{{\SetFigFont{12}{14.4}{\rmdefault}{\mddefault}{\updefault}{\color[rgb]{0,0,0}0.10}%
}}}}}
\put(526,-2311){\makebox(0,0)[lb]{\smash{{\SetFigFont{12}{14.4}{\rmdefault}{\mddefault}{\updefault}{\color[rgb]{0,0,0}$10.05^{\pm 0.10}$}%
}}}}
{\color[rgb]{0,0,0}\put(1330,-4855){\line( 5,-2){217.241}}
}%
{\color[rgb]{0,0,0}\put(1332,-4933){\line( 5,-2){217.241}}
}%
{\color[rgb]{0,0,0}\put(4801,-1411){\line( 0,-1){3900}}
}%
{\color[rgb]{0,0,0}\put(1801,-1411){\line( 0,-1){3900}}
}%
{\color[rgb]{0,0,0}\put(5082,-5234){\vector(-1, 0){300}}
}%
{\color[rgb]{0,0,0}\put(1520,-5239){\vector( 1, 0){300}}
}%
{\color[rgb]{0,0,0}\put(1801,-1486){\vector(-1, 0){  0}}
\put(1801,-1486){\vector( 1, 0){3000}}
}%
{\color[rgb]{0,0,0}\put(5026,-4636){\vector( 0,-1){  0}}
\put(5026,-4636){\vector( 0, 1){2400}}
}%
\put(1876,-1861){\makebox(0,0)[lb]{\smash{{\SetFigFont{12}{14.4}{\rmdefault}{\mddefault}{\updefault}{\color[rgb]{0,0,0}field cage barrel}%
}}}}
\put(1501,-4711){\rotatebox{90.0}{\makebox(0,0)[lb]{\smash{{\SetFigFont{12}{14.4}{\rmdefault}{\mddefault}{\updefault}{\color[rgb]{0,0,0}0.15}%
}}}}}
\put(5326,-4200){\rotatebox{90.0}{\makebox(0,0)[lb]{\smash{{\SetFigFont{12}{14.4}{\rmdefault}{\mddefault}{\updefault}{\color[rgb]{0,0,0}$\diameter = 720^{\, \pm 0.7}$}%
}}}}}
\put(3076,-1411){\makebox(0,0)[lb]{\smash{{\SetFigFont{12}{14.4}{\rmdefault}{\mddefault}{\updefault}{\color[rgb]{0,0,0}$610^{\, \pm 1}$}%
}}}}
\end{picture}%

 \caption{{\it{Mechanical accuracy specifications for the field cage: 
 The end flanges were required to be parallel with deviations less than $\unit[150]{\upmu m}$. The nominal axis of the field cage
 is defined as perpendicular to the anode end face in the center of the field cage. The measured axis of the field cage is specified
 to be within a tube with a diameter of $\unit[100]{\upmu m}$ with respect to the reference axis over the whole length of the field cage.
The distance of the first field strip to the anode end face was specified to be $\unit[10.05 \pm 0.10]{mm}$.}}}
 \label{accuracyrequirements}
 \end{figure}
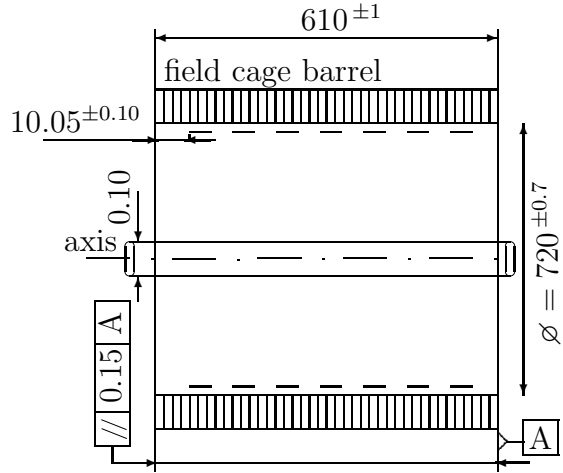

Detailed tolerance specifications for the field cage (Fig.~\ref{accuracyrequirements})
were derived from a study of field quality degradation due to an imperfect chamber geometry 
and the impact on the achievable point resolution~\cite{Schade:2009zz}.

Most critical is the correct alignment of the field cage axis relative to the anode end flanges. 
A misalignment of the axis produces a sheared field cage. This causes radial components of
the electric field which deteriorate the point resolution in the $r\varphi$ plane. Therefore
the tolerance on the alignment of the axis relative to the normal of the anode end face is defined 
most stringently to be within $\unit[100]{\upmu m}$.

Less critical is the parallel alignment of the anode relative to the cathode. 
A misalignment produces mainly field deviations along the $z$-axis
and to a lesser degree in the radial direction.
Hence the parallel alignment of the cathode relative to the anode was defined less stringently 
and required to be precise within $\unit[150]{\upmu m}$.

The length of the field cage is not a critical parameter because it can be adjusted by 
positioning the cathode inside the field cage. Therefore the specification has a comparably large tolerance of $\unit[1]{mm}$.
Similarly, the field cage diameter has a larger tolerance and is dimensioned to be $\unit[720.0 \pm 0.3]{mm}$.

\section{Production of the Field Cage Barrel}
\label{ProductionSection}
\begin{figure}[t!]
\centering
\subfigure[mandrel assembled with field strip board]{
\includegraphics[width = 0.45\textwidth]{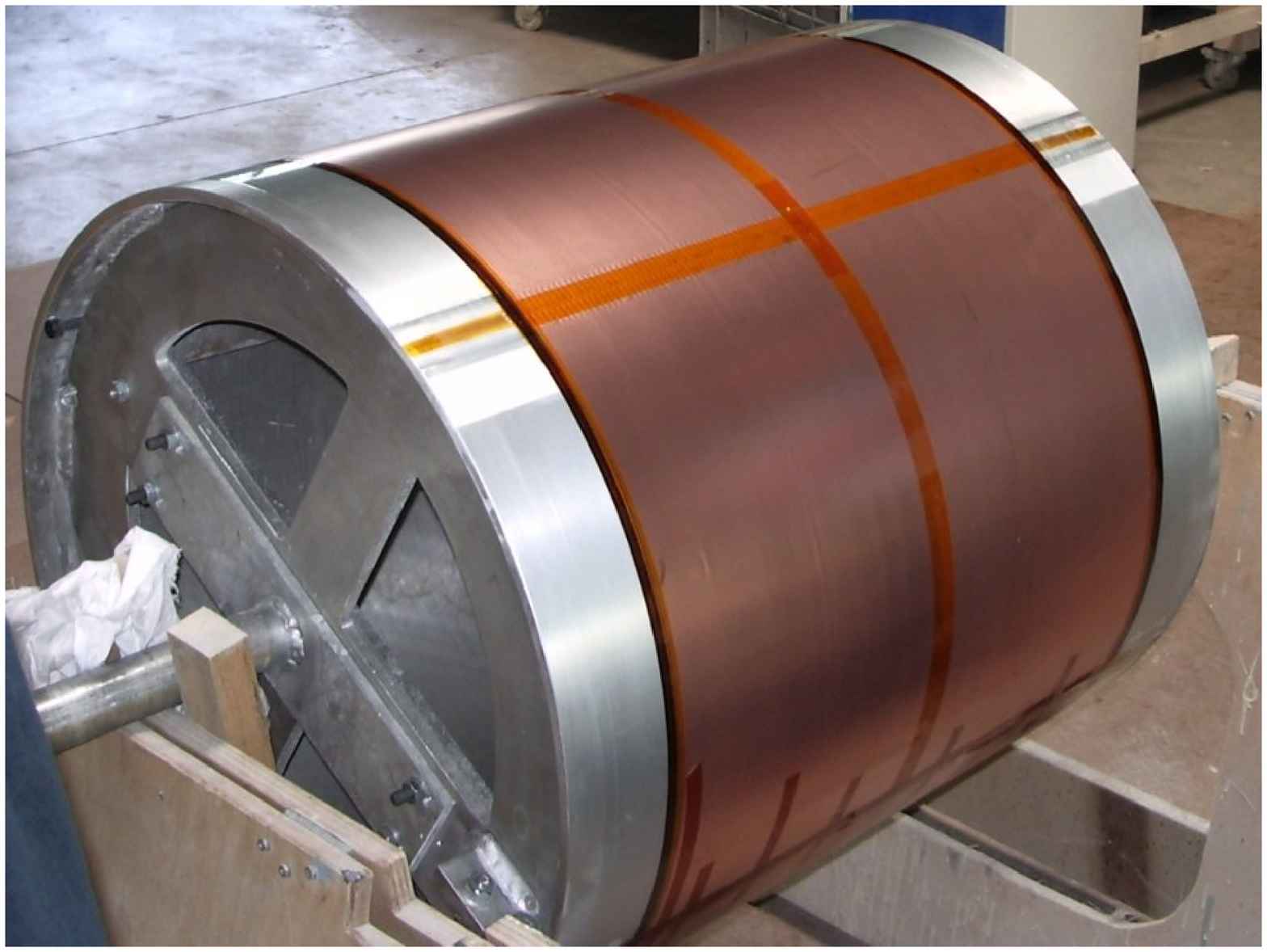}
\label{MandrellwithFoil}
}
\subfigure[lamination of the inner GRP layer]{
\includegraphics[width = 0.45\textwidth]{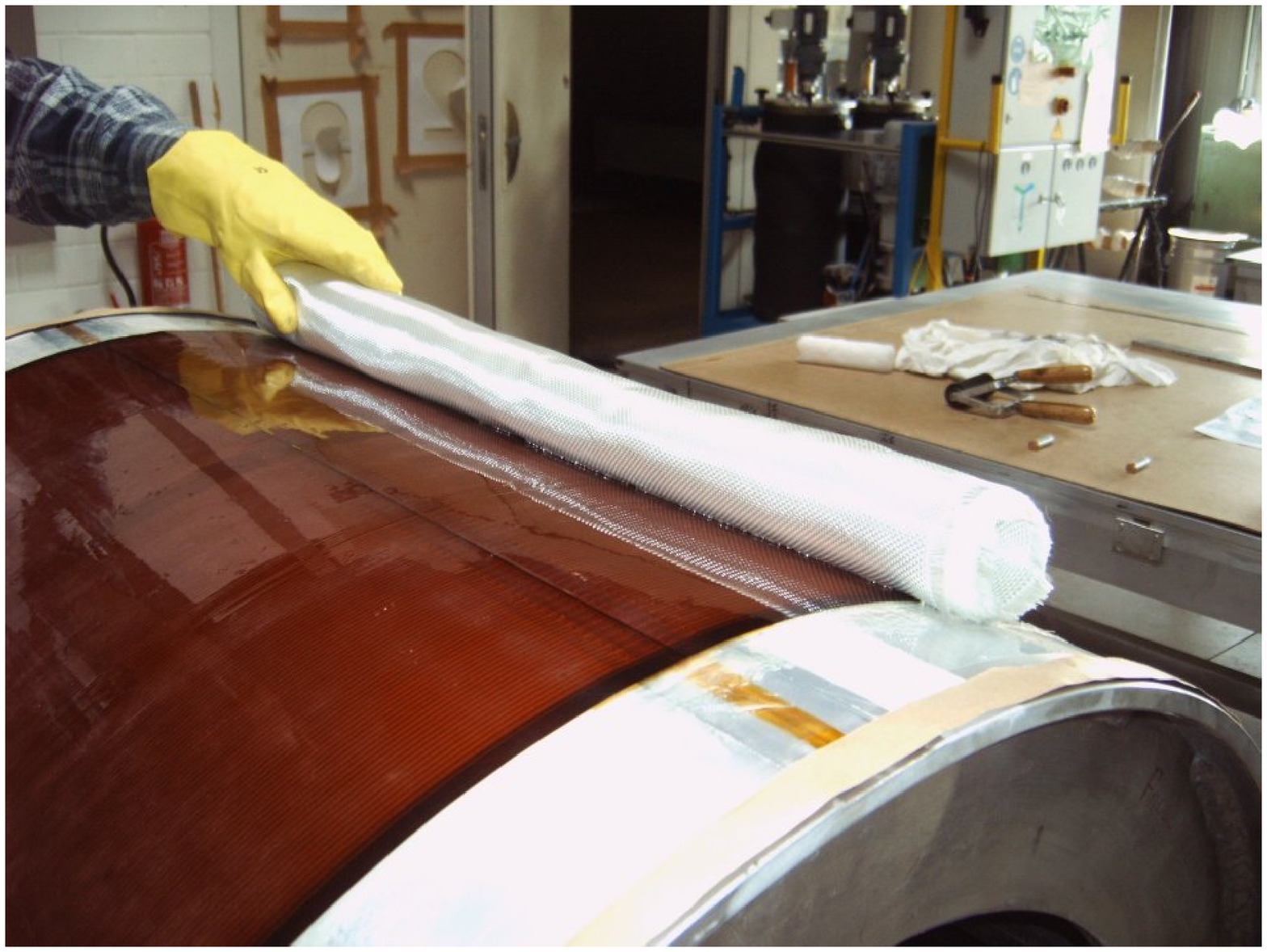}
\label{GRPlayer}
}
\caption{\it{Construction of the field cage on a mandrel.}}
\end{figure} 

The field cage was manufactured\footnote{DESY in cooperation with Haindl, individuelle Kunststoffverbundbauweise, \href{http://www.haindl-kunststoff.de}{http:/\hspace{-0.6mm}/www.haindl-kunststoff.de}} 
over a forming tool which served as a mold.
This was a 75-cm long mandrel with a diameter of $\unit[72]{cm}$ -- according to the field cage's
inner diameter. The mandrel could be reduced in diameter by a few millimeter via an expansion slot.

In the first step of the production, the field strip board was positioned on the mandrel
(Fig.~\ref{MandrellwithFoil}). Two 1-mm deep slots had been machined into the mandrel surface to accommodate the 
resistors on the field strip board. 
Then, the different layers of the field cage wall were laminated onto 
the foil. 
For the production of the GRP, first a glass-fiber canvas was put onto the mandrel (Fig.~\ref{GRPlayer}) and
moisturized with epoxy glue. Afterwards, air inclusions were removed from the layer with an underpressure treatment 
and the epoxy cured at $\unit[60]{^\circ C}$. The curing temperature was kept as low as possible to reduce 
thermal stresses on the field cage.

\begin{figure}[t!]
\subfigure[center points of circles fitted to reference points taken on the inside of the barrel]{
\includegraphics[width = 0.48\textwidth]{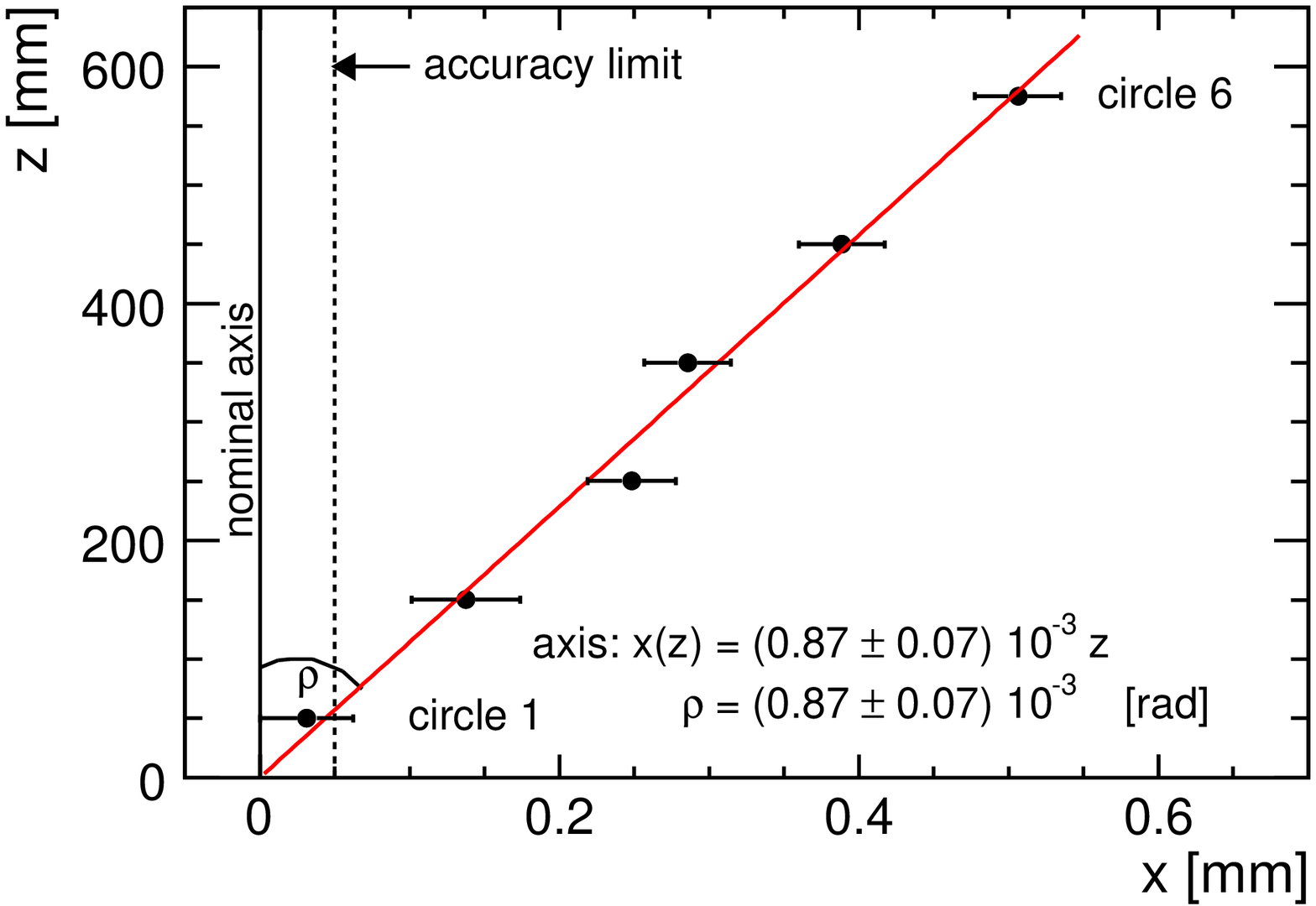}
\label{circlecenters}
}
\subfigure[distance of the first strip to the anode end face]{
\includegraphics[width = 0.48\textwidth]{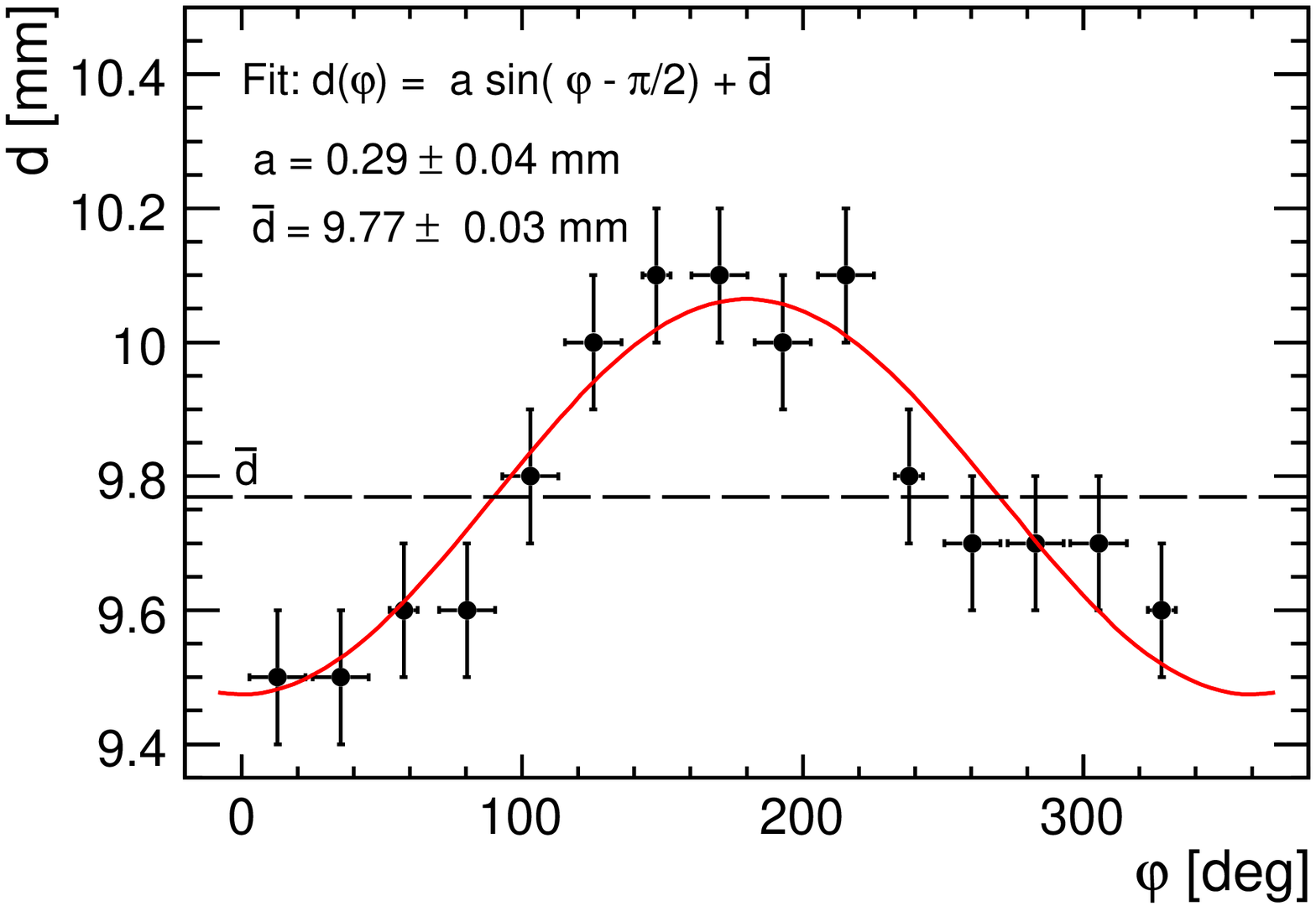}
\label{distancefirststrips}
}
\caption{\it{Determination of the field cage axis: The axis of the chamber is tilted and reaches an offset of 
$\unit[500]{\upmu m}$ at the cathode. 
The coordinates $z,r$ and $\varphi$ define a
cylindrical coordinate system for the field cage, with $z$ pointing in the direction of the nominal chamber axis, 
normal to the plane defined by the anode end face. $d$ is the distance of the first field
strip to the anode end face (Fig.~\ref{accuracyrequirements}).}}
\end{figure}

In the following steps of the production, the pre-produced end flanges and the honeycomb were 
laminated onto the inner GRP layer. On top, a layer of aramid paper sealed the cells of the honeycomb (see Fig.~\ref{SketchofWall}). 
A direct lamination of the outer GRP layer onto the open honeycomb could have filled the cells with epoxy and 
caused a higher and inhomogeneous material buildup of the wall.
The shielding layer of copper loaded polyimide completed the field cage.

With the lamination finished, the surfaces of the end flanges were machined for flatness and parallelism.
Finally the mandrel was reduced in diameter and removed from the field cage.

\subsection{Production Quality Assurance}
\label{ConstructionSurvey}
The important accuracy parameters for the field cage were surveyed in the commissioning phase of the LP at DESY. 
For this, about 100 measurement points were taken over the barrel with a spatial accuracy of $\unit[25]{\upmu m}$.

The end flanges of the field cage were
found to be parallel with deviations below $\unit[40]{\upmu m}$, while the
length of the field cage was measured to be $\unit[610.4 \pm 0.1]{mm}$. 
The diameter of the chamber was determined to $\unit[720.20 \pm 0.07]{mm}$ over the whole length of the barrel.
These numbers are in agreement with the specifications. 

To determine the axis of the field cage, measurement points were taken on 
the barrel inside at six fixed distances
relative to the anode reference plane. Each set of points defines a circle on the inside of the barrel and 
the center points of the six circles define the field cage axis.
A tilt of the axis was found, which results in a maximum offset of $\unit[500]{\upmu m}$ relative to the 
nominal position at the cathode (Fig.~\ref{circlecenters}). 
The angle between the measured axis and the nominal one
was determined to be $\rho = \unit[0.87 \pm 0.07]{mrad}$.

A second measurement of the axis was performed to confirm this result. 
For this, the distance $d$ of the first field strip to the anode end face was determined at several 
places around the circumference. 
The field strips on the inside define parallel planes perpendicular to the field cage axis.
Hence, $d$ has a fixed value if the axis is aligned correctly.

However, the measured distance $d$ varies sinus-like around the circumference 
(Fig.~\ref{distancefirststrips}).
The amplitude of the sinus is $\unit[0.3]{mm}$ and equal 
to $\rho \cdot r_\text{i}$. Here, $r_\text{i} = \unit[360]{mm}$ is the inner 
radius of the LP and $\rho$ the angle between the measured and the nominal field cage axis determined
with the initial method (Fig.~\ref{circlecenters}). 
Thus the amplitude has the expected magnitude, so that both methods agree on the misalignment of the axis.

\begin{figure}[t]
\centering

\setlength{\unitlength}{2800sp}%
\begingroup\makeatletter\ifx\SetFigFont\undefined%
\gdef\SetFigFont#1#2#3#4#5{%
  \reset@font\fontsize{#1}{#2pt}%
  \fontfamily{#3}\fontseries{#4}\fontshape{#5}%
  \selectfont}%
\fi\endgroup%
\begin{picture}(4074,4056)(1264,-5323)
\thinlines
{\color[rgb]{0,0,0}\put(1425,-5089){\line( 0,-1){150}}
\put(1425,-5239){\line( 1, 0){150}}
}%
{\color[rgb]{0,0,0}\put(1279,-4784){\line( 1, 0){296}}
}%
{\color[rgb]{0,0,0}\put(4880,-5049){\line( 1, 0){150}}
}%
{\color[rgb]{0,0,0}\multiput(4807,-4977)(6.25000,-6.25000){13}{\makebox(1.6667,11.6667){\SetFigFont{5}{6}{\rmdefault}{\mddefault}{\updefault}.}}
\multiput(4882,-5052)(-6.25000,-6.25000){13}{\makebox(1.6667,11.6667){\SetFigFont{5}{6}{\rmdefault}{\mddefault}{\updefault}.}}
}%
{\color[rgb]{0,0,0}\put(1279,-4185){\line( 1, 0){296}}
}%
{\color[rgb]{0,0,0}\put(5026,-5161){\framebox(300,300){}}
}%
{\color[rgb]{0,0,0}\put(1276,-5081){\framebox(300,1270){}}
}%
{\color[rgb]{0,0,0}\put(1809,-4578){\line( 18, -1){2995}}
}%
{\color[rgb]{0,0,0}\put(1797,-4802){\line( 18, -1){2995}}
}%
{\color[rgb]{0,0,0}\put(1810,-2161){\line( 18, -1){2994}}
}%
{\color[rgb]{0,0,0}\put(1801,-1936){\line( 18, -1){3000}}
}%
{\color[rgb]{0,0,0}\put(5101,-2911){\vector( 0,-1){490}}
}%
{\color[rgb]{0,0,0}\put(5101,-3886){\vector( 0, 1){305}}
}%
{\color[rgb]{0,0,0}\put(1811,-3362){\line( 1, 0){3270}}
}%
{\color[rgb]{0,0,0}\put(1811,-5236){\line( 1, 0){2990}}
}%
\put(5070,-5108){\makebox(0,0)[lb]{\smash{{\SetFigFont{12}{14.4}{\rmdefault}{\mddefault}{\updefault}{\color[rgb]{0,0,0}A}%
}}}}
\put(1501,-4111){\rotatebox{90.0}{\makebox(0,0)[lb]{\smash{{\SetFigFont{12}{14.4}{\rmdefault}{\mddefault}{\updefault}{\color[rgb]{0,0,0}A}%
}}}}}
{\color[rgb]{0,0,0}\put(1330,-4855){\line( 5,-2){217.241}}
}%
{\color[rgb]{0,0,0}\put(1332,-4933){\line( 5,-2){217.241}}
}%
{\color[rgb]{0,0,0}\put(4801,-1411){\line( 0,-1){3900}}
}%
{\color[rgb]{0,0,0}\put(1801,-1411){\line( 0,-1){3900}}
}%
{\color[rgb]{0,0,0}\put(5082,-5234){\vector(-1, 0){300}}
}%
{\color[rgb]{0,0,0}\put(1520,-5239){\vector( 1, 0){300}}
}%
{\color[rgb]{0,0,0}\put(1801,-1486){\vector(-1, 0){  0}}
\put(1801,-1486){\vector( 1, 0){3000}}
}%
{\color[rgb]{0,0,0}\put(4801,-1411){\line( 0,-1){3900}}
}%
{\color[rgb]{0,0,0}\put(1884,-4583){\line( 0,-1){223}}
}%
{\color[rgb]{0,0,0}\put(1958,-4586){\line( 0,-1){225}}
}%
{\color[rgb]{0,0,0}\put(2033,-4590){\line( 0,-1){224}}
}%
{\color[rgb]{0,0,0}\put(2108,-4593){\line( 0,-1){225}}
}%
{\color[rgb]{0,0,0}\put(4504,-4719){\line( 0,-1){224}}
}%
{\color[rgb]{0,0,0}\put(4579,-4724){\line( 0,-1){225}}
}%
{\color[rgb]{0,0,0}\put(4654,-4727){\line( 0,-1){225}}
}%
{\color[rgb]{0,0,0}\put(4729,-4731){\line( 0,-1){224}}
}%
{\color[rgb]{0,0,0}\put(1801,-1411){\line( 0,-1){3900}}
}%
{\color[rgb]{0,0,0}\put(2194,-1957){\line( 0,-1){225}}
}%
{\color[rgb]{0,0,0}\put(2269,-1960){\line( 0,-1){225}}
}%
{\color[rgb]{0,0,0}\put(2344,-1963){\line( 0,-1){225}}
}%
{\color[rgb]{0,0,0}\put(2419,-1969){\line( 0,-1){224}}
}%
{\color[rgb]{0,0,0}\put(2494,-1972){\line( 0,-1){225}}
}%
{\color[rgb]{0,0,0}\put(2569,-1975){\line( 0,-1){225}}
}%
{\color[rgb]{0,0,0}\put(2644,-1980){\line( 0,-1){223}}
}%
{\color[rgb]{0,0,0}\put(2719,-1984){\line( 0,-1){225}}
}%
{\color[rgb]{0,0,0}\put(2794,-1987){\line( 0,-1){225}}
}%
{\color[rgb]{0,0,0}\put(2869,-1991){\line( 0,-1){224}}
}%
{\color[rgb]{0,0,0}\put(2944,-1996){\line( 0,-1){225}}
}%
{\color[rgb]{0,0,0}\put(3019,-1999){\line( 0,-1){225}}
}%
{\color[rgb]{0,0,0}\put(3168,-2008){\line( 0,-1){225}}
}%
{\color[rgb]{0,0,0}\put(3243,-2011){\line( 0,-1){225}}
}%
{\color[rgb]{0,0,0}\put(3318,-2014){\line( 0,-1){225}}
}%
{\color[rgb]{0,0,0}\put(3393,-2020){\line( 0,-1){225}}
}%
{\color[rgb]{0,0,0}\put(3468,-2023){\line( 0,-1){225}}
}%
{\color[rgb]{0,0,0}\put(3543,-2026){\line( 0,-1){225}}
}%
{\color[rgb]{0,0,0}\put(3618,-2031){\line( 0,-1){223}}
}%
{\color[rgb]{0,0,0}\put(3693,-2035){\line( 0,-1){225}}
}%
{\color[rgb]{0,0,0}\put(3768,-2038){\line( 0,-1){225}}
}%
{\color[rgb]{0,0,0}\put(3843,-2042){\line( 0,-1){224}}
}%
{\color[rgb]{0,0,0}\put(3918,-2047){\line( 0,-1){225}}
}%
{\color[rgb]{0,0,0}\put(3993,-2050){\line( 0,-1){225}}
}%
{\color[rgb]{0,0,0}\put(4068,-2054){\line( 0,-1){224}}
}%
{\color[rgb]{0,0,0}\put(4142,-2059){\line( 0,-1){225}}
}%
{\color[rgb]{0,0,0}\put(4217,-2062){\line( 0,-1){225}}
}%
{\color[rgb]{0,0,0}\put(4292,-2065){\line( 0,-1){225}}
}%
{\color[rgb]{0,0,0}\put(4367,-2071){\line( 0,-1){225}}
}%
{\color[rgb]{0,0,0}\put(4442,-2074){\line( 0,-1){225}}
}%
{\color[rgb]{0,0,0}\put(2119,-1952){\line( 0,-1){224}}
}%
{\color[rgb]{0,0,0}\put(2045,-1948){\line( 0,-1){225}}
}%
{\color[rgb]{0,0,0}\put(1970,-1945){\line( 0,-1){225}}
}%
{\color[rgb]{0,0,0}\put(1896,-1940){\line( 0,-1){224}}
}%
{\color[rgb]{0,0,0}\put(4516,-2077){\line( 0,-1){225}}
}%
{\color[rgb]{0,0,0}\put(4591,-2082){\line( 0,-1){223}}
}%
{\color[rgb]{0,0,0}\put(4666,-2086){\line( 0,-1){225}}
}%
{\color[rgb]{0,0,0}\put(4741,-2089){\line( 0,-1){225}}
}%
{\color[rgb]{0,0,0}\put(2183,-4598){\line( 0,-1){225}}
}%
{\color[rgb]{0,0,0}\put(2258,-4601){\line( 0,-1){225}}
}%
{\color[rgb]{0,0,0}\put(2333,-4605){\line( 0,-1){225}}
}%
{\color[rgb]{0,0,0}\put(2408,-4610){\line( 0,-1){225}}
}%
{\color[rgb]{0,0,0}\put(2483,-4613){\line( 0,-1){225}}
}%
{\color[rgb]{0,0,0}\put(2558,-4617){\line( 0,-1){224}}
}%
{\color[rgb]{0,0,0}\put(2633,-4622){\line( 0,-1){224}}
}%
{\color[rgb]{0,0,0}\put(2708,-4625){\line( 0,-1){225}}
}%
{\color[rgb]{0,0,0}\put(2781,-4629){\line( 0,-1){224}}
}%
{\color[rgb]{0,0,0}\put(2856,-4634){\line( 0,-1){224}}
}%
{\color[rgb]{0,0,0}\put(2931,-4637){\line( 0,-1){225}}
}%
{\color[rgb]{0,0,0}\put(3006,-4641){\line( 0,-1){224}}
}%
{\color[rgb]{0,0,0}\put(3081,-4644){\line( 0,-1){225}}
}%
{\color[rgb]{0,0,0}\put(3156,-4649){\line( 0,-1){225}}
}%
{\color[rgb]{0,0,0}\put(3231,-4652){\line( 0,-1){225}}
}%
{\color[rgb]{0,0,0}\put(3306,-4656){\line( 0,-1){225}}
}%
{\color[rgb]{0,0,0}\put(3381,-4661){\line( 0,-1){225}}
}%
{\color[rgb]{0,0,0}\put(3456,-4664){\line( 0,-1){225}}
}%
{\color[rgb]{0,0,0}\put(3531,-4668){\line( 0,-1){224}}
}%
{\color[rgb]{0,0,0}\put(3606,-4673){\line( 0,-1){224}}
}%
{\color[rgb]{0,0,0}\put(3681,-4676){\line( 0,-1){225}}
}%
{\color[rgb]{0,0,0}\put(3756,-4680){\line( 0,-1){224}}
}%
{\color[rgb]{0,0,0}\put(3831,-4685){\line( 0,-1){224}}
}%
{\color[rgb]{0,0,0}\put(3906,-4688){\line( 0,-1){225}}
}%
{\color[rgb]{0,0,0}\put(3981,-4692){\line( 0,-1){224}}
}%
{\color[rgb]{0,0,0}\put(4056,-4695){\line( 0,-1){225}}
}%
{\color[rgb]{0,0,0}\put(4129,-4700){\line( 0,-1){225}}
}%
{\color[rgb]{0,0,0}\put(4204,-4703){\line( 0,-1){225}}
}%
{\color[rgb]{0,0,0}\put(4279,-4707){\line( 0,-1){225}}
}%
{\color[rgb]{0,0,0}\put(4354,-4712){\line( 0,-1){225}}
}%
{\color[rgb]{0,0,0}\put(4429,-4715){\line( 0,-1){225}}
}%
{\color[rgb]{0,0,0}\put(5082,-5234){\vector(-1, 0){300}}
}%
{\color[rgb]{0,0,0}\put(1520,-5239){\vector( 1, 0){300}}
}%
{\color[rgb]{0,0,0}\put(1801,-1486){\vector(-1, 0){  0}}
\put(1801,-1486){\vector( 1, 0){3000}}
}%
{\color[rgb]{0,0,0}\put(3094,-2003){\line( 0,-1){224}}
}%
{\color[rgb]{0,0,0}\put(1726,-4561){\vector( 0,-1){  0}}
\put(1726,-4561){\vector( 0, 1){2400}}
}%
{\color[rgb]{0,0,0}\put(1806,-3360){\line( 18, -1){451}}
}%
{\color[rgb]{0,0,0}\put(2699,-3422){\line( 18, -1){451}}
}%
{\color[rgb]{0,0,0}\put(4505,-3542){\line( 18, -1){451}}
}%
{\color[rgb]{0,0,0}\put(3598,-3481){\line( 18, -1){451}}
}%
\put(1876,-1861){\makebox(0,0)[lb]{\smash{{\SetFigFont{12}{14.4}{\rmdefault}{\mddefault}{\updefault}{\color[rgb]{0,0,0}field cage barrel}%
}}}}
\put(1501,-4711){\rotatebox{90.0}{\makebox(0,0)[lb]{\smash{{\SetFigFont{12}{14.4}{\rmdefault}{\mddefault}{\updefault}{\color[rgb]{0,0,0}0.04}%
}}}}}
\put(1876,-1861){\makebox(0,0)[lb]{\smash{{\SetFigFont{12}{14.4}{\rmdefault}{\mddefault}{\updefault}{\color[rgb]{0,0,0}field cage barrel}%
}}}}
\put(5026,-3136){\rotatebox{90.0}{\makebox(0,0)[lb]{\smash{{\SetFigFont{12}{14.4}{\rmdefault}{\mddefault}{\updefault}{\color[rgb]{0,0,0}0.5}%
}}}}}
\put(4148,-3676){\rotatebox{356.0}{\makebox(0,0)[lb]{\smash{{\SetFigFont{12}{14.4}{\rmdefault}{\mddefault}{\updefault}{\color[rgb]{0,0,0}axis}%
}}}}}
\put(3226,-3286){\makebox(0,0)[lb]{\smash{{\SetFigFont{12}{14.4}{\rmdefault}{\mddefault}{\updefault}{\color[rgb]{0,0,0}nominal axis}%
}}}}
\put(1651,-3661){\rotatebox{90.0}{\makebox(0,0)[lb]{\smash{{\SetFigFont{12}{14.4}{\rmdefault}{\mddefault}{\updefault}{\color[rgb]{0,0,0}$\diameter = 720.2^{\,  \pm 0.07}$}%
}}}}}
\put(2851,-1411){\makebox(0,0)[lb]{\smash{{\SetFigFont{12}{14.4}{\rmdefault}{\mddefault}{\updefault}{\color[rgb]{0,0,0}$610.4^{\, \pm 0.1}$}%
}}}}
\end{picture}%

\caption{\it{Measured shape of the LP: 
The requirements in length, alignment of the end flanges and roundness of the barrel are fulfilled, 
but alignment of the field cage axis does not satisfy the accuracy goal.}}
\label{Fieldcageshape}
\end{figure}
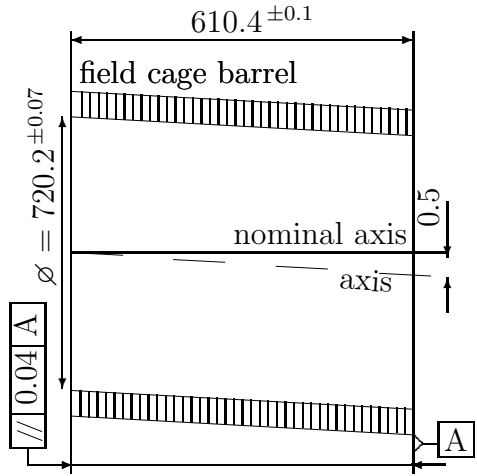
\begin{figure}
\centering
\includegraphics[width = 0.48\textwidth]{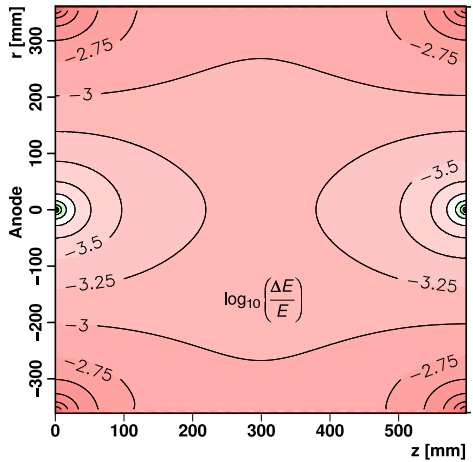}
\caption{\it{Calculated field quality: 
Due to the shear of the field cage (see Fig.~\ref{Fieldcageshape}), the calculated electric field
inside the LP is homogeneous only to a level of $\Delta E/E \approx 10^{-3}$.}}
\label{rechnung}
\end{figure}

Figure~\ref{Fieldcageshape} illustrates the measured shape of the field cage. 
Due to the shear of the barrel, the electric drift field inside the chamber is not homogeneous to the required level  
(Fig.~\ref{rechnung}). The field inhomogeneities have a magnitude of $10^{-4} \le \Delta E/E \lesssim 10^{-3}$.
 
\section{Extrapolation to the ILD TPC}
\label{extrapolationToILD}
For the ILD, a TPC is planned with a diameter of the inner field cage of $\unit[65]{cm}$, 
of the outer field cage of $\unit[360]{cm}$ and a drift distance of $\unit[215]{cm}$. This is about 3.5 times longer than the LP.
At the same time, the magnetic field of ILD is $\unit[3.5]{T}$ compared to $\unit[1]{T}$ for the LP.
As mentioned in section \ref{introsection}, the ratio $L/B$ of the magnetic field to the drift distance $L$ 
is the same for both TPCs and so are the required relative mechanical accuracy specifications. 

Scaling the mechanical tolerances of the LP by a factor of three yields a tolerance for the alignment of the field cage axis
in the range of $\unit[300]{\upmu m}$ and a required parallel alignment of anode and cathode of $\unit[450]{\upmu m}$ for the ILD TPC. 

The main challenge for the design of the ILD TPC will be the reduction of the material budget of the wall 
to $\unit[1\%]{X_0}$ while increasing the high voltage stability to ${\cal{O}}(\unit[100]{kV})$.

Starting from the current LP wall cross section (see Fig.~\ref{SketchofWall}),
a reduction of the material budget is possible by thinning down the field strips
to $\unit[20]{\upmu m}$ and by replacing copper by aluminum. 
In addition, with further optimization studies of the chamber statics and mechanical tests, the  
thickness of the GRP could be diminished. This would reduce the contribution of epoxy and glass-fiber to the material budget.
Assuming a moderate optimization, GRP layers of $\unit[200]{\upmu m}$ could be sufficiently stable to construct
a self supporting tube of $\unit[4.3]{m}$ length for the inner field cage.

The LP wall samples were tested to be high voltage stable up to at least $\unit[30]{kV}$. 
In the wall sample tested, a single polyimide layer of $\unit[50]{\upmu m}$ 
was introduced which can withstand $\unit[10]{kV}$ alone.
The insulating honeycomb-GRP structure increased the high voltage stability to above $\unit[30]{kV}$. 

\begin{figure}[t]
\centering
\includegraphics[width=0.7\textwidth]{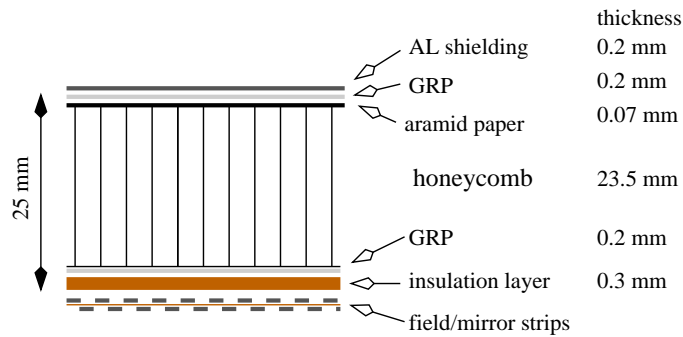}
\caption{\it{First draft of the cross section for the wall of the inner field cage of the ILD TPC.}}
\label{SketchoffinalWall}
\end{figure}

Extrapolating to the ILD TPC, the wall of the inner field cage could have a cross section as shown in Figure~\ref{SketchoffinalWall}.
Here, an insulation which is equivalent to a single $\unit[300]{\upmu m}$ thick polyimide layer together with 
the honeycomb sandwich provide a high voltage stability in the range of $\unit[70]{kV}$.
This wall has a material budget of $\unit[1\%]{X_0}$, which is the design value.
However, the detailed fabrication of the thicker polyimide layer still has to be evaluated and tested.
\enlargethispage{1ex}

The outer field cage of the ILD TPC will be a single barrel structure serving as gas vessel and high voltage insulation. 
Its material budget goal is planned to be $\unit[2\%]{X_0}$ at most.
At the same time the wall must be thicker than the one for the inner field cage to gain sufficient mechanical robustness. 
A wall thickness of $\unit[60]{mm}$, which could provide a sufficient stability, can be realized by scaling up the thickness 
of the honeycomb material and doubling the thickness of the GRP layers.
In this case, the material budget would reach the design value of $\unit[2\%]{X_0}$. 

It must be stated, that the mechanical
and the high voltage stability, both for the proposed inner and outer field cage wall, need to be quantified 
by dedicated calculations and sample piece tests. Also the precise mechanical accuracy specifications have to be 
revisited on the basis of further studies, also taking into account the final detector gas.

\section*{Summary}
The LP is the first TPC prototype with a size relevant for a TPC of a future ILC detector. 
The length of the LP is $\unit[61]{cm}$ and the inner diameter of the field cage barrel of $\unit[72]{cm}$ 
is similar to the inner field cage for the ILD TPC. 

The design of the chamber was optimized for a high electric field homogeneity of 
$\Delta E/E \lesssim 10^{-4}$ and a low material budget of the walls of $\unit[1.21\%]{X_0}$. This is 
close to the final design value of $\unit[1\%]{X_0}$. Further optimizations of the 
wall structure are under study and the final design goal of $\unit[1\%]{X_0}$ per wall seems to be in reach.

The LP is part of a test beam infrastructure which is installed at the 6-GeV DESY electron test beam. 
This infrastructure was realized in the framework of the \mbox{EUDET} project~\cite{EUDET}
and became available in November 2008. Since then it is in use by different research groups doing
R\&D work for a TPC of detector at a future linear collider \cite{LCTPC}.

\section*{Acknowledgments}
This work is supported by the Commission of the European Communities under the 6$^\text{th}$ Framework Programme 
`Structuring the European Research Area', contract number RII3-026126.
We thank the whole LCTPC collaboration for sharing their expertise with us in the design phase of the Large Prototype field cage.
The Department of Physics of the University of Hamburg provided a valuable technical support, 
in particular B. Frensche, U. Pelz and the mechanical workshop.
We thank the Technical University of Hamburg-Harburg, especially P. G\"{u}hrs, 
for the collaboration in performing the mechanical sample piece tests 
and the advice in the construction of the field cage.

\bibliographystyle{hieeetr.bst}

\end{document}